\documentclass[prb,aps,showpacs,superscriptaddress,twocolumn]{revtex4}
\usepackage{latexsym}
\usepackage{amsfonts}
\usepackage{amssymb}
\usepackage{amsmath}
\usepackage{graphicx}
\begin{document}
\title{Macroscopic evidence for quantum criticality and field-induced \\
quantum fluctuations in cuprate superconductors}
\date{\today}
\author{A. D. Beyer,$^1$ V. S. Zapf,$^2$ H. Yang,$^1$, F. Fabris,$^2$
M. S. Park,$^3$ K. H. Kim,$^3$ S.-I. Lee,$^3$ and N.-C. Yeh}

\affiliation{Department of Physics, California Institute of
Technology, Pasadena, CA \\ $^2$National High Magnetic Field
Laboratory, Los Alamos, NM \\ $^3$Department of Physics, Pohang
University of Science and Technology, Pohang, Korea}

\begin{abstract}
We present {\it macroscopic} experimental evidence for field-induced
{\it microscopic} quantum fluctuations in different hole- and electron-type
cuprate superconductors with varying doping levels and numbers of CuO$_2$ layers
per unit cell. The significant suppression of the zero-temperature in-plane
magnetic irreversibility field relative to the paramagnetic field
in all cuprate superconductors suggests strong quantum fluctuations due to
the proximity of the cuprates to quantum criticality.
\end{abstract}
 \maketitle

High-temperature superconducting cuprates are extreme type-II superconductors that
exhibit strong thermal, disorder and quantum fluctuations in their vortex
states.~\cite{FisherDS91,Blatter94,Yeh93,Balents94,Giamarchi95,Kotliar96,Kierfeld04,Zapf05,Yeh05a}
While much research has focused on the {\it macroscopic} vortex dynamics of cuprate superconductors
with phenomenological descriptions,~\cite{FisherDS91,Blatter94,Yeh93,Balents94,Giamarchi95,Kierfeld04}
little effort has been made to address the {\it microscopic} physical origin of their extreme
type-II nature.~\cite{Yeh05a} Given that competing orders (CO) can exist in the ground
state of these  doped Mott insulators besides superconductivity
(SC),~\cite{Yeh05a,Zhang97,Demler01,Chakravarty01,Sachdev03,Kivelson03,LeePA06}
the occurrence of quantum criticality may be expected.~\cite{Demler01,Sachdev03,Onufrieva04}
The proximity to quantum criticality and the existence of CO can significantly
affect the low-energy excitations of the cuprates due to strong quantum
fluctuations~\cite{Zapf05,Yeh05a} and the redistribution of quasiparticle spectral weight
among SC and CO.~\cite{Yeh05a,ChenCT07,Beyer06} Indeed, empirically the low-energy excitations
of cuprate superconductors appear to be unconventional, exhibiting intriguing
properties unaccounted for by conventional Bogoliubov quasiparticles.~\cite{Yeh05a,ChenCT07,Beyer06,ChenCT03}
Moreover, external variables such as temperature ($T$) and applied magnetic field
($H$) can vary the interplay of SC and CO, such as inducing or enhancing the
CO~\cite{ChenHY05a,Lake01} at the price of more rapid
suppression of SC, thereby leading to weakened superconducting stiffness and strong
thermal and field-induced fluctuations.~\cite{FisherDS91,Blatter94,Yeh93}
On the other hand, the quasi two-dimensional nature of the cuprates can also
lead to quantum criticality in the limit of decoupling of CuO$_2$ planes.
In this work we demonstrate experimental evidence from {\it macroscopic}
magnetization measurements for field-induced quantum fluctuations among
a wide variety of cuprate superconductors with different {\it microscopic}
variables such as the doping level ($\delta$) of holes or electrons, and the
number of CuO$_2$ layers per unit cell ($n$).~\cite{Chakravarty04} We suggest that
the manifestation of strong field-induced quantum fluctuations is consistent
with a scenario that all cuprates are in close proximity to a quantum critical
point (QCP).~\cite{Kotliar96}

To investigate the effect of quantum fluctuations on the vortex dynamics
of cuprate superconductors, our strategy involves studying the vortex phase diagram
at $T \to 0$ to minimize the effect of thermal fluctuations, and applying
magnetic field {\it parallel} to the CuO$_2$ planes ($H \parallel ab$) to minimize
the effect of random point disorder. The rationale for having $H \parallel ab$
is that the intrinsic pinning effect of layered CuO$_2$ planes generally dominates
over the pinning effects of random point disorder, so that the commonly
observed glassy vortex phases associated with point disorder for
$H \parallel c$ ({\it e.g.} vortex glass and Bragg glass)~\cite{FisherDS91,Giamarchi95,Kierfeld04}
can be prevented. In the {\it absence} of quantum fluctuations, random point disorder
can cooperate with the intrinsic pinning effect to stabilize the low-temperature
vortex smectic and vortex solid phases,~\cite{Balents94}
so that the vortex phase diagram for $H \parallel ab$ would
resemble that of the vortex-glass and vortex-liquid phases observed for
$H \parallel c$ with a glass transition $H_G (T = 0)$ approaching
$H_{c2} (T = 0)$. On the other hand, when {\it field-induced quantum fluctuations}
are dominant, the vortex phase diagram for $H \parallel ab$ will deviate
substantially from predictions solely based on thermal fluctuations
and intrinsic pinning, and we expect strong suppression of the magnetic
irreversibility field $H_{irr} ^{ab}$ relative to the upper critical field
$H_{c2} ^{ab}$ at $T \to 0$, because the induced persistent current circulating
along both the c-axis and the ab-plane can no longer be sustained if
field-induced quantum fluctuations become too strong to maintain the c-axis
superconducting phase coherence.

In this communication we present experimental results that are consistent with the notion that
all cuprate superconductors exhibit significant field-induced quantum fluctuations
as manifested by a characteristic field $H_{irr} ^{ab} (T \to 0) \equiv H^{\ast} \ll H_{c2} ^{ab} (T \to 0)$.
Moreover, we find that we can express the degree of quantum fluctuations for each cuprate in terms of
a reduced field $h^{\ast} \equiv H^{\ast}/H_{c2}^{ab}(0)$,
with $h^{\ast} \to 0$ indicating strong quantum fluctuations and
$h^{\ast} \to 1$ referring to the mean-field limit. Most important,
the $h^{\ast}$ values of all cuprates appear to follow a trend
on a $h^{\ast} (\alpha)$-vs.-$\alpha$ plot, where $\alpha$ is a material parameter
for a given cuprate that reflects its doping level, electronic anisotropy,
and charge imbalance if the number of CuO$_2$ layers per unit cell $n$ satisfies
$n \ge 3$.~\cite{Kotegawa01a,Kotegawa01b} In the event that
$H_{c2} ^{ab} (0)$ exceeds the paramagnetic field
$H_p \equiv \Delta _{\rm SC} (0)/(\sqrt{2} \mu _B)$ for highly anisotropic
cuprates, where $\Delta _{\rm SC} (0)$ denotes the superconducting gap
at $T = 0$, $h^{\ast}$ is defined by $(H^{\ast}/H_p)$
because $H_p$ becomes the maximum critical field for superconductivity.

To find $h^{\ast}$, we need to determine both the upper critical field
$H_{c2} ^{ab} (T)$ and the irreversibility field $H_{irr} ^{ab} (T)$ to as
low temperature as possible. Empirically, $H_{c2} ^{ab} (T)$ can be derived
from measuring the magnetic penetration depth in pulsed fields, with $H_{c2}^{ab}(0)$
extrapolated from $H_{c2} ^{ab} (T)$ values obtained at finite temperatures.
The experiments involve measuring the frequency shift $\Delta f$ of a tunnel
diode oscillator (TDO) resonant tank circuit with the sample
contained in one of the component inductors.
Details of the measurement techniques have been given in Ref.~\onlinecite{Zapf05}.
In general we find that the condition $H_{c2} ^{ab} (0) > H_p$ is satisfied
among all samples investigated so that we define $h^{\ast} \equiv (H^{\ast}/H_p)$
hereafter. On the other hand, determination of $H_{c2}^{ab} (0)$ and $H_{c2}^c (0)$ 
is still useful because it provides the electronic anisotropy
$\gamma \equiv (\xi _{ab}/\xi _c) = \lbrack H_{c2}^{ab}(0)/H_{c2}^c(0) \rbrack$,
where $\xi _{ab} (\xi _c)$ refers to the in-plane (c-axis) superconducting coherence length.

\begin{figure}
  \centering
  \includegraphics[width=3.375in]{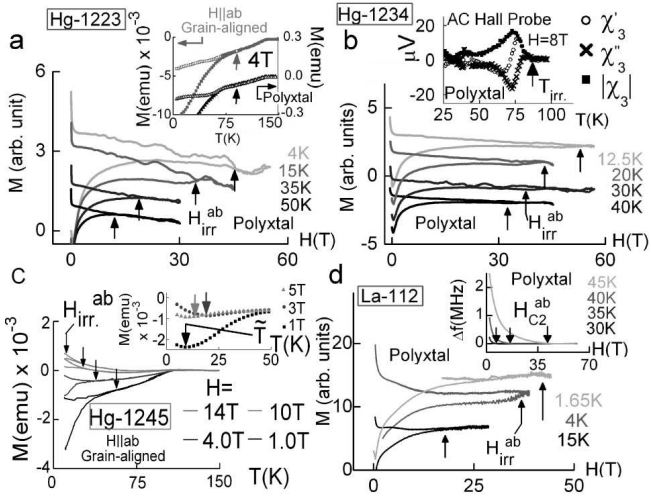}
\caption{Representative measurements of the in-plane irreversibility fields
$H_{irr}^{ab} (T)$ in cuprate superconductors: (a) Hg-1223 (polycrystalline
and grain-aligned), (b) Hg-1234 (polycrystalline), (c) Hg-1245 (grain-aligned),
and (d) La-112 (polycrystalline and grain-aligned). Insets: (a) Consistent
$T_{irr}^{ab}(H)$ obtained from maximum irreversibility of a
polycrystalline sample and from irreversibility of a grain-aligned
sample with $H \parallel ab$; (b) representative $\chi_3$ data taken using AC
Hall probe techniques; (c) details of the $M$-vs.-$T$ curves,
showing an anomalous upturn at $T < \tilde T$; (d) exemplifying determination
of $H_{c2}^{ab}$ in La-112 using TDO to measure $\Delta f$.\cite{Zapf05}}
\label{fig1}
\end{figure}

To determine $H_{irr} ^{ab} (T)$, we employed different experimental techniques,
including DC measurements of the magnetization $M(T,H)$ with the use of a SQUID
magnetometer or a homemade Hall probe magnetometer for lower fields (up to 9 Tesla),
a DC magnetometer (up to 14 Tesla) at the National
High Magnetic Field Laboratory (NHMFL) in Los Alamos (LANL-PPMS), a cantilever magnetometer
at the NHMFL in Tallahasseefor higher fields (up to 33 Tesla DC fields in a $^3$He 
refrigerator),~\cite{Naughton97} and a compensated coil for magnetization measurements 
in the pulsed-field (PF) facilities at LANL for even higher fields
(up to 65 Tesla pulsed fields in a $^3$He refrigerator).~\cite{Zapf05} 
In addition, AC measurements of the third harmonic magnetic susceptibility ($\chi _3$) 
as a function of temperature in a constant field are employed to determine the onset 
of non-linearity in the low-excitation limit.~\cite{ReedDS95} Examples of the 
measurements of $H_{irr} ^{ab} (T)$ for $\rm HgBa_2Ca_2Cu_3O_x$
(Hg-1223, $T_c = 133$ K), $\rm HgBa_2Ca_3Cu_4O_x$ (Hg-1234, $T_c = 125$ K),
$\rm HgBa_2Ca_4Cu_5O_x$ (Hg-1245, $T_c = 108$ K) and $\rm Sr_{0.9}La_{0.1}CuO_2$
(La-112, $T_c = 43$ K), are shown in Fig.~1 (a) - (d), and the consistency among
$H_{irr} ^{ab} (T)$ deduced from different techniques have been verified,
as summarized in the $H$-vs.-$T$ phase diagrams ($H \parallel ab$) in Fig.~2(a) -- (d).
The Hg-based cuprates are in either polycrystalline or grain-aligned forms,
and the quality of these samples is confirmed with x-ray diffraction
and magnetization measurements to ensure single phase and nearly 100\% volume
superconductivity.~\cite{KimMS98,IyoA06} We have also
verified that $H_{irr}^{ab} (T)$ obtained from the polycrystalline
samples are consistent with those derived from the grain-aligned samples with
$H \parallel ab$, because the measured irreversibility in a polycrystalline
sample is manifested by its maximum irreversibility $H_{irr}^{ab}$
among grains of varying orientation relative to the applied field.
Examples of this consistency have been shown in Ref.~\onlinecite{Zapf05}
and also in the main panel and the inset of Fig.~1(a).

In addition to the four different cuprates considered in this work, we
compare measurements of $H_{irr} ^{ab} (T)$ on other cuprate superconducting single
crystals, including underdoped $\rm YBa_2Cu_3O_{7-\delta}$ (Y-123, $T_c$ = 87 K),~\cite{OBrien00}
optimally doped $\rm Nd_{1.85}Ce_{0.15}CuO_4$ (NCCO, $T_c$ = 23 K)~\cite{Yeh92}
and overdoped $\rm Bi_2Sr_2CaCu_2O_x$ (Bi-2212, $T_c$ = 60 K).~\cite{Krusin-Elbaum04}
The irreversibility fields for these cuprates normalized to their
corresponding paramagnetic fields $H_p$ are summarized in Fig.~3(a) as a
function of the reduced temperature $(T/T_c)$, clearly demonstrating
strong suppression of $H^{\ast}$ relative to $H_p$ and $H_{c2} ^{ab} (0)$
in all cuprates and implying significant field-induced quantum fluctuations.

\begin{figure}
  \centering
  \includegraphics[width=3.45in]{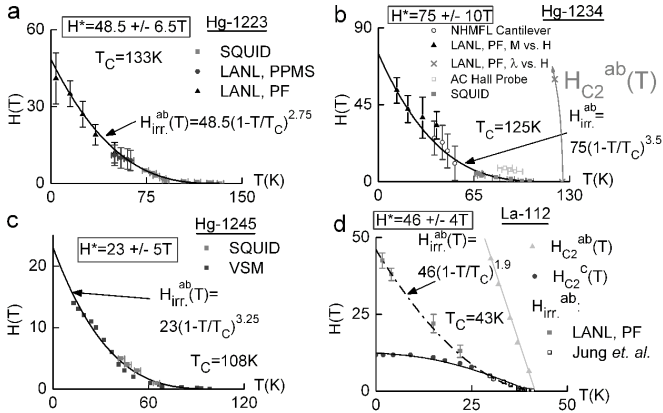}
\caption{Determination of $H^{\ast}$ using various
magnetic measurements of $H_{irr}^{ab} (T)$:
(a) Hg-1223, (b) Hg-1234, (c) Hg-1245, and (d) La-112. In (b) and (d)
dashed lines indicate $H_{c2}^{ab} (T)$ from TDO measurements.
In (d) we also illustrate $H_{c2}^{c}$ for comparison. We note reasonable
consistency among different experimental techniques, indicating strong suppression of
$H^{\ast} \equiv H_{irr} ^{ab} (0)$ relative to $H_{c2} ^{ab} (0)$ (or $H_p$)
in all cuprates.}
\label{fig2}
\end{figure}

\begin{table*}
\caption{Quantum criticality parameters among different cuprates. All fields in tesla. $\sigma$ denotes a parameter's uncertainty.}
\begin{ruledtabular}
\begin{tabular}{|c|c|c|c|c|c|c|c|c|c|c|c|c|c|}
                    & \boldmath{$\delta$}   &   \boldmath{$\delta_{o}$} & \boldmath{$\delta_{i}$}   & \boldmath{$\gamma$}                                  & \boldmath{$\sigma_{\gamma}$}  &   \boldmath{$\alpha(10^{-2})$}            &   \boldmath{$\sigma_{\alpha}(10^{-3})$}           & \boldmath{$H^{\ast}$}       & \boldmath{$\sigma_{H^{\ast}}$}    & \boldmath{$H_{c2}^{ab}[H_{P}]$}      & \boldmath{$\sigma_{H_{P}}$}  &   \boldmath{$h^{\ast}$}    & \boldmath{$\sigma_{h^{\ast}}$}   \\
\hline Hg-1245      &       $0.15$          &       $1.30$              &       $0.80$              &       $55$~\cite{Hg1245comment}                       &       $25$                    &        $0.06$                             &       $0.3$                                       &       $23.0$                &   $5.0$                           &   $-[278]$                           &       $40$                            &       $0.08$            &       $0.02$                  \\
\hline Hg-1223      &       $0.15$          &       $1.04$              &       $0.92$              &       $52$~\cite{Zech96}                              &       $18$                    &        $0.26$                             &       $0.9$                                       &       $48.5$                &   $6.5$                           &   $-[347]$                           &       $50$                            &       $0.14$            &       $0.02$                  \\
\hline Hg-1234      &       $0.15$          &       $1.20$              &       $0.80$              &       $52$~\cite{Zech96}                              &       $10$                    &        $0.13$                             &       $0.2 $                                      &       $75.0$                &   $10.0$                           &   $-[320]$                           &       $46$                            &       $0.23$            &       $0.02$                  \\
\hline La-112       &       $0.10$          &       $1.00$              &       $1.00$              &       $13$~\cite{Zapf05}                              &       $4.0$                   &        $0.77$                             &       $2.4 $                                      &       $46.0$                &   $4.0$                           &   $160[110]$                         &       $10$                            &       $0.42$            &       $0.04$                  \\
\hline Bi-2212      &       $0.225$         &       $1.00$              &       $1.00$              &       $11$~\cite{Krusin-Elbaum04}                     &       $8.0$                   &        $2.05$                             &       $15  $                                      &       $65.0$                &   $10$                            &   $100[155]$                         &       $22$                            &       $0.42$            &       $0.06$                  \\
\hline NCCO         &       $0.15$          &       $1.00$              &       $1.00$              &       $13$~\cite{Yeh92}                               &       $5.0$                   &        $1.15$                             &       $4.4 $                                      &       $40.0$                &   $5.0$                           &   $77[59]$                         &       $8.0$                           &       $0.68$            &       $0.12$                  \\
\hline Y-123        &       $0.13$          &       $1.00$              &       $1.00$              &       $7.0$~\cite{Zech96}                             &       $2.0$                   &        $1.86$                             &       $5.3 $                                      &       $210$                 &   $50$                            &   $600[239]$                         &       $25$                            &       $0.88$            &       $0.23$                  \\
\end{tabular}
\end{ruledtabular}
\label{table1}
\end{table*}

The physical significance of $h^{\ast}$ may be better understood by considering how the
magnetic irreversibility for $H \parallel ab$ occurs. For sufficiently low $T$ and small $H$,
a supercurrent circulating both along and perpendicular to the CuO$_2$ planes with a coherent
superconducting phase can be induced and sustained, leading to magnetic irreversibility.
On the other hand, strong thermal or quantum fluctuations due to large anisotropy and
competing orders in the cuprates can reduce the phase coherence of supercurrents, 
particularly the coherence perpendicular to the CuO$_2$ planes, thereby diminishing 
the magnetic irreversibility.
Thus, we expect that the degree of the in-plane magnetic irreversibility is dependent
on the nominal doping level $\delta$, the electronic anisotropy $\gamma$,
the number of CuO$_2$ layers per unit cell $n$, and the ratio of charge imbalance
$(\delta _o/\delta _i)$~\cite{Kotegawa01a,Kotegawa01b} between the doping level of the
outer layers ($\delta _o$) and that of the inner layer(s) ($\delta _i$) in multi-layer cuprates
with $n \ge 3$. In other words, $h^{\ast}$ for each cuprate superconductor
may be expressed in terms of a material parameter $\alpha$ that depends on $\delta$,
$\gamma$, $n$ and $(\delta _o / \delta _i)$, and the simplest assumption for a
linearized dependence of $\alpha$ on these variables gives:
\begin{eqnarray}
\alpha &\equiv \gamma ^{-1} \delta (\delta _o/\delta _i)^{-(n-2)}, \qquad (n \ge 3); \\
\alpha &\equiv \gamma ^{-1} \delta , \qquad \qquad \qquad \qquad (n \le 2) .
\label{eq:alpha}
\end{eqnarray}
If the suppression of the in-plane magnetic irreversibility is associated with
field-induced quantum fluctuations and the proximity to a quantum critical
point $\alpha _c$,~\cite{Demler01} $h^{\ast} (\alpha)$ should be a function
of $|\alpha - \alpha _c|$. Indeed, we find that using the
empirically determined values for different cuprates tabulated in Table 1 and the
definition of $\alpha$ given above, the $h^{\ast}$-vs.-$\alpha$ data for a wide
variety of cuprates appear to follow a trend, as shown in Fig.~3(b).
For comparison, we include in Fig.~3(b) theoretical curves predicted for
field-induced static spin density waves (SDW) in cuprate superconductors in
the limit of $H_{c1} (0) \ll H \ll H_{c2} (0)$, where $h^{\ast}$ (above which static
\begin{figure}[h]
  \centering
  \includegraphics[width=3.45in]{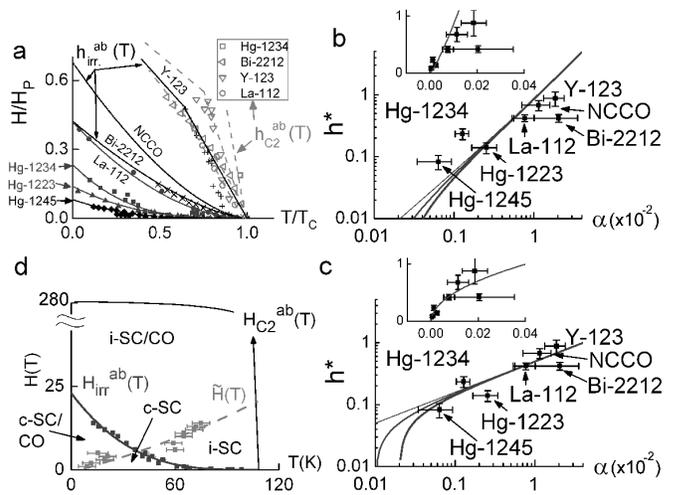}
\caption{(a) Reduced in-plane fields $(H_{irr}^{ab}/H_p)$
and $(H_{c2}^{ab}/H_p)$ vs. $(T/T_c)$ for various cuprates.
In the $T \to 0$ limit where $H_{irr}^{ab} \to H^{\ast}$, the reduced
fields $h^{\ast} \equiv (H^{\ast}/H_p) < 1$ for all cuprates are listed
in Table I for Y-123, NCCO, Bi-2212, La-112, Hg-1234, Hg-1223, and Hg-1245 (in descending order).
(b) $h^{\ast}$ vs. $\alpha$ in logarithmic plot for different
cuprates, with decreasing $\alpha$ representing increasing quantum fluctuations.
The lines given by $-400 |\alpha - \alpha _c| / \ln | \alpha - \alpha _c |$
represent the field-induced SDW scenario~\cite{Demler01} in Eq.~(3) with different
$\alpha _c$ = 0, $10^{-4}$ and $2 \times 10^{-4}$ from left to right.
Inset: The linear plot of the main panel. (c) The same data as in (b) are
compared with the power-law dependence (solid lines) given
by $5(\alpha - \alpha _c)^{1/2}$, using different $\alpha _c$ = 0, $10^{-4}$
and $2 \times 10^{-4}$ from left to right. Inset: The linear plot
of the main panel. (d) The $H$-vs.-$T$ diagram of Hg-1245. (See text for details).}
\label{fig3}
\end{figure}
SDW coexists with SC) satisfies the relation:~\cite{Demler01}
\begin{equation}
h^{\ast} (\alpha) \propto |\alpha - \alpha _c|/\lbrack \ln |\alpha - \alpha _c| \rbrack.
\label{eq:hstar}
\end{equation}
Here $\alpha _c$ is a non-universal critical point,~\cite{Demler01}
$h^{\ast} (\alpha) \to 0$ for $\alpha \to \alpha _c$, and we have shown theoretical
curves associated with three different $\alpha_c$ values for comparison with data.
On the other hand, a simple scaling argument would assert a power-law dependence:
\begin{equation}
h^{\ast} (\alpha) \propto |\alpha - \alpha _c|^{a}. \qquad \qquad (a > 0)
\label{eq:hpower}
\end{equation}
Using $a \sim 0.5$ in Eq.~(4), we compare the power-law dependence with experimental data
in Fig.~3(c). This dependence appears to agree better with experimental data than the
SDW/SC formalism in Eq.~(3).

Although our available data cannot accurately determine $\alpha _c$, we further
examine Hg-1245 (which has the smallest $h^{\ast}$) for additional clues associated with
the nature of the QCP. We find that the magnetization $M$ of Hg-1245 always exhibits
an anomalous increase for $T < \tilde T (H)$ (see the inset of Fig.~1(c)), indicating
a field-induced reentry of magnetic ordering below $\tilde T (H)$. This magnetism reentry
line $\tilde H (T)$ is shown together with $H_{irr}^{ab} (T)$ in Fig.~3(d).
We suggest that the regime below both $H_{irr}^{ab}$ and $\tilde H$ 
corresponds to a coherent SC state ($c$-SC), and that bounded by $H_{irr}^{ab}$ and
$\tilde H$ is associated with a coherent phase of coexisting SC
and magnetic CO ($c$-SC/CO), whereas that above $H_{irr}^{ab}$ is an incoherent SC phase
($i$-SC and $i$-SC/CO) with strong fluctuations.

Our conjecture of a field-induced magnetic CO in Hg-1245 contributing to 
quantum fluctuations may be further corroborated by considering 
the $h^{\ast}$-vs.-$\alpha$ dependence in the multi-layered cuprates
Hg-1223, Hg-1234 and Hg-1245. While these cuprate superconductors have the
highest $T_c$ and $H_{c2}$ values, as shown in Table I, they also exhibit the smallest
$h^{\ast}$ and $\alpha$ values, suggesting maximum quantum fluctuations. These strong
quantum fluctuations can be attributed to both their extreme two dimensionality 
(i.e., large $\gamma$)~\cite{KimMS98,KimMS01} and significant charge imbalance 
that leads to strong CO in the inner layers.~\cite{Kotegawa01a,Kotegawa01b} Indeed 
muon spin resonance ($\mu$SR) experiments~\cite{Tokiwa03} have revealed increasing 
antiferromagnetic ordering in the inner layers of the multi-layer cuprates 
with $n \ge 3$. Given that the $\gamma$ values of all Hg-based multi-layer cuprates
are comparable (Table I), the finding of larger quantum fluctuations (i.e. smaller 
$h^{\ast}$) in Hg-1245 is suggestive of increasing quantum fluctuations with
stronger competing order. However, further investigation of the $h^{\ast}$ and $\gamma$
values of other multi-layer cuprates will be necessary to confirm whether 
competing orders in addition to large anisotropy contribute to quantum fluctuations. 

In summary, our investigation of the {\it in-plane} magnetic irreversibility
in a wide variety of cuprate superconductors reveals strong field-induced quantum
fluctuations. The {\it macroscopic} irreversibility field exhibits
dependences on such {\it microscopic} material parameters as the doping level,
the charge imbalance in multi-layered cuprates, and the electronic anisotropy.
Our finding is consistent with the notion that cuprate superconductors
are in close proximity to quantum criticality.

\begin{acknowledgments}
Research at Caltech was supported by NSF Grant DMR-0405088 and through
the NHMFL. The SQUID data were taken at the Beckman Institute at Caltech.
Work at Pohang University was supported by the Ministry of Science and Technology
of Korea. The authors gratefully acknowledge Dr. Kazuyasu Tokiwa and
Dr. Tsuneo Watanabe at the Tokyo University of Science for providing the
$\rm HgBa_2Ca_4Cu_5O_x$ (Hg-1245) samples.
\end{acknowledgments}


\begin{thebibliography}{35}
\bibitem[{Hg1()}]{Hg1245comment}
\bibinfo{note}{The $\gamma$ value for Hg-1245 was extrapolated from those of 
Hg-1201, Hg-1212, Hg-1223 and Hg-1234~\cite{Zech96}, with the latter three
being comparable within experimental errors.}
\end{thebibliography}

\end{document}